\documentclass[journal,10pt]{IEEEtran}
\usepackage{mathrsfs}
\usepackage{amsfonts}
\usepackage{graphicx,cite,epsfig,amssymb,amsmath}
\usepackage{color,xcolor}
\definecolor{red}{rgb}{1.00, 0.00, 0.00}
\usepackage{pifont}
\usepackage{stmaryrd}
\usepackage{setspace}
\usepackage{subfigure}
\usepackage{cite}
\usepackage{array}
\usepackage{float}
\usepackage{multirow}
\usepackage{algorithm,algpseudocode}
\usepackage{epstopdf}
\usepackage{mathtools}
\usepackage{diagbox}
\usepackage{booktabs}
\providecommand{\algorithmname}{Algorithm}
\floatname{algorithm}{\protect\algorithmname}
\newcommand{\bm}[1]{\mbox{\boldmath{$#1$}}}
\newcommand{\tabincell}[2]{\begin{tabular}{@{}#1@{}}#2\end{tabular}}
\usepackage{arydshln}
\begin{document}
	
\title{Learning to Branch: Accelerating Resource Allocation in Wireless Networks}
\author{Mengyuan Lee, Guanding Yu, \emph{Senior Member, IEEE}, and Geoffrey Ye Li  \emph{Fellow, IEEE}
	\thanks{Copyright (c) 2015 IEEE. Personal use of this material is permitted. However, permission to use this material for any other purposes must be obtained from the IEEE by sending a request to pubs-permissions@ieee.org.}
\thanks{This work was supported by the National Key Research and Development Program of China under Grant 2018YFB1802302 and by the Fundamental Research Funds for the Central Universities. This paper was presented in part at IEEE VTC 2019-Fall, Honolulu, HI, USA, Sep. 2019.}
	\thanks{M. Lee and  G. Yu are with the Zhejiang Provincial Key Laboratory of Information Processing, Communication and Networking, Zhejiang University, Hangzhou 310027, China. e-mail: \{mengyuan\_lee, yuguanding\}@zju.edu.cn.}
	\thanks{G. Y. Li is with the School of ECE, Georgia Institute of Technology, Atlanta, GA, USA. e-mail: liye@ece.gatech.edu.}
}
\maketitle

\begin{abstract}
Resource allocation in wireless networks, such as device-to-device (D2D) communications, is usually formulated as mixed integer nonlinear programming (MINLP) problems, which are generally NP-hard and difficult to get the optimal solutions. Traditional methods to solve these MINLP problems are all based on  mathematical optimization techniques, such as the branch-and-bound (B\&B) algorithm that converges slowly and has forbidding complexity for real-time implementation. Therefore, machine leaning (ML) has been used recently to address the MINLP problems in wireless communications. In this paper, we use imitation learning method to accelerate the B\&B algorithm. With invariant problem-independent features and appropriate problem-dependent feature selection for D2D communications, a good auxiliary prune policy can be learned in a supervised manner to speed up the most time-consuming branch process of the B\&B algorithm. Moreover, we develop a mixed training strategy to further reinforce the generalization ability and a deep neural network (DNN) with a novel loss function to achieve better dynamic control over optimality and computational complexity. Extensive simulation demonstrates that the proposed method can achieve good optimality and reduce computational complexity simultaneously.
\end{abstract}


%%%% ---------------------- \input{sub/introduce.tex} ----------------------

\section{Introduction}
To meet with the increasing demand for higher data rate and better QoS requirements, device-to-device (D2D) communications have been proposed as a promising technique for the LTE-Advanced networks and have recently attracted significant attention from both academic and industry communities \cite{d2d1,d2d2,d2d3,d2d5,d2d6}. The key idea of D2D communications is enabling proximity users to communicate directly by reusing the wireless spectrum of conventional cellular users (CUs). Compared with traditional cellular communications, this new technique can significantly reduce the traffic load in the core network and increase the network throughput as well. However, D2D communications may generate severe interference to the existing CUs due to spectrum reusing. Proper resource allocation can improve the system performance while reducing the co-channel interference between D2D and CUs.

Resource allocation for D2D communications and other wireless networks is usually formulated as mixed integer nonlinear programming (MINLP) problems, which are in general NP-hard and no efficient global optimal algorithm is available yet. Existing solutions to those MINLP problems are usually based on various mathematical optimization techniques. In general, the optimal solutions to the MINLP problems can only be achieved by the branch-and-bound (B\&B) algorithm  \cite{bb}. However, the worst-case computational complexity of the B\&B algorithm is exponential, causing it impractical for real-time implementation. Therefore, some works aim to develop near-optimal or sub-optimal algorithms and reduce computational complexity, such as methods based on game theory \cite{gametheory1,gametheory2} or graph theory \cite{graph1,graph2}. Heuristic algorithms, which alternatively or iteratively update the combinatorial and continuous variables, have also been widely investigated \cite{heuristic1,heuristic2}. The sub-optimal methods usually suffer from two major shortcomings. On the one hand, the performance gaps between the sub-optimal solutions and the optimal ones are hard to control. On the other hand, many iterative based heuristic algorithms still have high computational complexity for real-time implementation although they are faster than the B\&B algorithm.

The aforementioned two disadvantages can be overcome by ML techniques. Its inference stage is generally very fast and we can adopt different techniques to improve and control the output performance. Therefore, this paper incorporates the machine learning (ML) technique and mathematical optimization technique to address the resource allocation in D2D communications, which can be also applied to resource allocation in other wireless networks. We are inspired by recently emerged studies on applying the ML techniques to address mathematical optimization problems in wireless communications \cite{shi,mlop3,guo,li1,review,yuwei,wcl}. In \cite{shi,mlop3}, deep neural networks (DNNs) have been utilized to deal with the power control problems in wireless networks. In \cite{guo,li1,review}, reinforcement learning has been adopted to solve resource allocation problems for various wireless networks. In \cite{yuwei}, spatial learning has been developed to schedule interfering links in D2D networks without the need of channel state information (CSI). In \cite{wcl},  linear sum assignment problems (LSAPs), which are often encountered in wireless communications, have been solved based on DNNs.  All the above studies follow the end-to-end learning paradigm \cite{shi} and treat a given resource optimization problem as a ``black box" to learn its input/output relation by various ML techniques. This end-to-end paradigm is very suitable for optimization problems with only one kind of output variables. For examples, problems in \cite{shi,mlop3,guo} only include continuous output variables, whereas \cite{yuwei,wcl} only involve combinatorial ones. However, such end-to-end paradigm is hard to effectively solve the MINLP problems due to the more complicated algorithm structures.

Therefore, this paper exploits the specific algorithm structures to solve the MINLP problems for resource allocation in D2D communications using ML technique. Specifically, we formulate a mathematical problem to maximize the minimum data rate among D2D pairs by jointly optimizing the channel allocation and power control. Instead of directly using mathematical techniques to solve it, we turn to a hybrid ML and mathematical method. We propose to leverage the imitation learning to reduce the computational complexity of the B\&B algorithm by accelerating the most time-consuming branch process. The imitation learning has been first proposed in \cite{Hhe} for the mixed integer linear programming (MILP) problems. Although this very approach has been introduced in wireless communications in \cite{LORA} to solve the resource allocation problems in cloud radio access networks (Cloud-RANs), there is no work applying the imitation learning to solve D2D resource allocation problems.

Imitation learning can be further converted into a binary classification problem with proper feature design, which can be solved by the classical support vector machine (SVM) in a supervised manner. To improve the learning performance, we carefully select the problem-dependent features for the resource allocation problems in D2D communications in addition to the problem-independent features in \cite{Hhe}. The problem-dependent features include the power constraint of each D2D link as well as the CSI of both D2D and interference links. We further investigate the influence of the problem-dependent features in terms of optimality and computational complexity.

The ML based algorithm is expected to have few needed training samples, fast training speed, and good generalization ability. Therefore, we make the following three major improvements on the specific learning algorithm as compared with the work in \cite{Hhe,LORA}. First, we simplify the learning task without performance loss by learning an auxiliary prune policy to prune the non-optimal nodes that are not fathomed. This improvement leads to fewer needed training samples and faster training speed than the existing one that discards the original prune policy in the B\&B algorithm and learns a new one to prune all the non-optimal nodes. Second, we propose to only collect the nodes that are visited under current policy and are not fathomed by the original B\&B prune policy to further accelerate the training process. This modification can also mitigate the dataset imbalance issue and improve the accuracy of the binary classifier. Third, we develop a mixed training strategy to reinforce the generalization ability of the imitation learning method. A novel loss function is also developed when utilizing the DNN as the binary classifier to dynamically control the optimality and computational complexity. By extensive simulation results, we verify the effectiveness of the proposed method for D2D communications and highlight some insights about practical implementation as well.

The rest of this paper is organized as follows. In Section II, we formulate resource allocation into an MINLP problem. In Section III, we will transform the problem into a more tractable one and develop the optimal B\&B algorithm to solve it. Section IV introduces the imitation learning method to accelerate the B\&B algorithm. In Section V, we present test results of the  proposed method, which motivate us to make two further improvements in Section VI. Finally, we will conclude this paper in Section VII.
%%%%% -------------------------- \input{sub/model.tex} -------------------------------
\section{Problem Formulation}
In this section, we formulate the resource allocation in wireless networks into an MINLP problem using D2D communications as an example.

\subsection{Resource Allocation in D2D Communications}
\begin{figure}[h]
	\centering
	\includegraphics[width=0.8\linewidth, height=0.18\textheight]{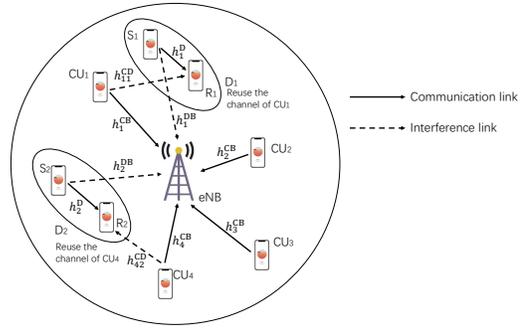}
	\caption{System model.}
	\label{fig:sysmodel}
\end{figure}

As depicted in Fig. \ref{fig:sysmodel}, we consider an uplink single-cell system with $K$ CUs in a set $\mathscr{K}=\{1, ..., K\}$ and $L$ D2D pairs in a set $\mathscr{L}=\{1, ..., L\}$. We assume that each uplink CU connects to the Evolved NodeB (eNB) with an orthogonal channel. Moreover, we assume that D2D pairs transmit data by reusing the uplink channels of CUs. In the D2D networks, the number of D2D pairs is usually smaller than that of cellular users \cite{number1,number2}. Therefore, we assume $K \geq L$ in this paper. Nevertheless, our proposed method can also be utilized while $K \leq L$.

As in Fig. \ref{fig:sysmodel}, we denote $h_{kl}^\text{CD}$ as the instantaneous channel power gain of the interference link between CU $k$ and the receiver of D2D pair $l$, $h_{k}^\text{CB}$ as the channel power gain between CU $k$ and the eNB, $h_l^\text{D}$ as the channel power gain between D2D pair $l$, and $h_{l}^\text{DB}$ as the channel power gain of the interference link  between the transmitter of D2D pair $l$ and the eNB.

We further introduce $\bm{\rho}= [\rho_{kl}]$ as the indicator vector of the channel allocation. Specifically, $\rho_{kl}=1$ if the channel of CU $k$ is reused by D2D pair $l$, and $\rho_{kl}=0$ otherwise. Denote $\bm{p^\text{C}}=[p_k^\text{C}]$ and  $\bm{p^\text{D}}=[p_{kl}^\text{D}]$ as the transmit power vectors for $K$ CUs and $L$ D2D pairs, respectively, where $p_k^\text{C}$ denotes the allocated transmit power for CU $k$ and $p_{kl}^\text{D}$ denotes the power of D2D pair $l$ on the channel of CU $k$.

Since each CU channel is assumed to be reused by at most one D2D pair, the signal-to-interference-plus-noise ratio (SINR)  of D2D pair $l$ on the channel of CU $k$ can be written as
$$SINR_{kl}^\text{D}(\bm{p^\text{C}},\bm{p^\text{D}},\bm{\rho})=\frac{\rho_{kl}p_{kl}^\text{D}h_l^\text{D}}{\sigma_N^2+p_k^\text{C}h_{kl}^\text{CD}},$$
where $\sigma_N^2$ denotes the power of the additive white Gaussian noise (AWGN). Similarly, the SINR achieved by CU $k$ can be expressed as
$$SINR_{k}^\text{C}(\bm{p^\text{C}},\bm{p^\text{D}},\bm{\rho})=\frac{p_k^\text{C}h_{k}^\text{CB}}{\sigma_N^2+\sum_{l\in \mathscr{L}} \rho_{kl}p_{kl}^\text{D}h_{l}^\text{DB}}.$$

Accordingly, the data rates in bits per second per hertz (i.e., normalized by the channel bandwidth) of CU $k$  and D2D pair $l$ on all channels can be written as
\begin{eqnarray}
R_{k}^\text{C}(\bm{p^\text{C}},\bm{p^\text{D}},\bm{\rho})=  \log(1+SINR_{k}^\text{C}(\bm{p^\text{C}},\bm{p^\text{D}},\bm{\rho})), \label{crate}
\end{eqnarray}
\begin{equation}
\begin{aligned}
R_{l}^\text{D}(\bm{p^\text{C}},\bm{p^\text{D}},\bm{\rho})&= \sum_{k\in \mathscr{K}}\rho_{kl} R_{kl}^\text{D}(\bm{p^\text{C}},\bm{p^\text{D}},\bm{\rho})\\ &= \sum_{k\in \mathscr{K}}\rho_{kl} \log(1+SINR_{kl}^\text{D}(\bm{p^\text{C}},\bm{p^\text{D}},\bm{\rho})),
\end{aligned}
\label{drate}
\end{equation}
respectively, where $R_{kl}^\text{D}$ is the data rate of D2D pair $l$ on the channel of CU $k$.

Resource allocation in D2D communications is to determine $\bm{p^\text{C}$, $p^\text{D}}$, and $\bm{\rho}$ to optimize the overall network performance.

\subsection{Problem Formulation}
To achieve fairness among different users, we specifically consider a resource allocation problem to maximize the minimum data rate of D2D pairs in this paper with the following constraints. First, the minimum data rate of each CU is required to be no less than given thresholds. Second, the power of individual links is also constrained. Finally,  as mentioned above, each channel can be reused by at most one D2D pair to limit the interference between different D2D pairs. Therefore, the resource allocation problem can be mathematically formulated as
\begin{eqnarray*}\label{Mresource}
	\hspace{5.4em}\max_{\{\bm{p^\text{C}}, \bm{p^\text{D}},\bm{\rho}\}} \min_{l\in \mathscr{L} } R_{l}^\text{D}(\bm{p^\text{C}},\bm{p^\text{D}},\bm{\rho}),\hspace{6em}
	\eqref{Mresource_obj} \label{Mresource_obj}
\end{eqnarray*}
\begin{subequations}
	subject to
	\begin{align}
	\rho_{kl} \in \{0,1\}, \quad{\forall}k\in \mathscr{K}, l\in \mathscr{L},\label{resource_sub1}
	\end{align}
	\begin{align}
	\sum_{l\in \mathscr{L}}\rho_{kl}\leq 1, \quad{\forall}k\in \mathscr{K},\label{resource_sub2}
	\end{align}
	\begin{align}
	\sum_{k\in \mathscr{K}}\rho_{kl}p_{kl}^\text{D} \leq P^\text{D}_{\max}, \quad{\forall}l\in \mathscr{L},\label{resource_sub3}	
	\end{align}
	\begin{align}
	R_{k}^\text{C}(\bm{p^\text{C}},\bm{p^\text{D}},\bm{\rho})\geq R^\text{C}_{\min}, \quad{\forall}k\in \mathscr{K},\label{resource_sub4}
	\end{align}
	\begin{align}
	p_{k}^\text{C}\leq P^\text{C}_{\max}, \quad{\forall}k\in \mathscr{K},\label{resource_sub5}
	\end{align}
\end{subequations}
where $R^\text{C}_{\min}$ is the minimum rate of CUs, $P^\text{C}_{\max}$ and $P^\text{D}_{\max}$ are the maximum transmit power level of CUs and D2D pairs, respectively. Note that, the proposed algorithm can be extended to other resource allocation problems with different objective functions  since the B\&B algorithm is a general approach to solve MINLP problems.

Problem  (\ref{Mresource}) is a typical resource allocation problem in D2D communications. Similar resource allocation problems can be also found in other wireless networks. It is an MINLP problem, which is NP-hard in general. The B\&B algorithm can find its optimal solution but with high computational complexity \cite{bb}. In the next section, we will focus on developing the B\&B algorithm for Problem (\ref{Mresource}) as the preparation for aforementioned ML-based acceleration process.

%%%%%% ------------------- \input{sub/decomp.tex} -------------------
\section{Optimal Branch-and-Bound Algorithm}
The B\&B algorithm is widely used for MINLP problems. However, most existing studies on resource allocation have focused on heuristic algorithms and only used the B\&B algorithm as the baseline without detailed development due to its high computational complexity. In this paper, we will accelerate the B\&B algorithm with the help of an ML technique, thus developing the optimal B\&B algorithm in detail is necessary. Therefore, in this section, we will first transform Problem (\ref{Mresource}) into a more tractable one and then develop the optimal B\&B algorithm to solve it.

\subsection{Problem Transformation}
To transform Problem  (\ref{Mresource}) into a more tractable one, we first derive a proposition about the optimal power allocation of D2D pair $l$ on the channel of CU $k$ in the following, which is proved in Appendix A.

\emph{Proposition 1:} If D2D pair $l$ is allowed to reuse the channel of CU $k$, then the optimal power of CU $k$ can be written as $$p_k^\text{C}=\frac{(2^{R_{\min}^\text{C}}-1)(\sigma_N^2+p_{kl}^\text{D}h_l^{\text{DB}})}{h_k^\text{CB}},$$ where  $p_{kl}^\text{D} \in [0, p_{kl}^{\max}]$ and $$p_{kl}^{\max}=\min\{(1/h_l^\text{DB})(p_{\max}^\text{C}h_k^\text{CB}/(2^{R_{\min}^\text{C}}-1)-\sigma_N^2), P_{\max}^\text{D}\}.$$

Based on Proposition 1 and the derivation in Appendix B, we can rewrite the data rate of D2D pair $l$ on all channels as a function of  $\{\bm{p^\text{D}},\bm{\rho}\}$  as stated in the following.

\emph{Proposition 2:} The data rate of D2D pair $l$ on all channels  defined in (\ref{drate}) can be rewritten into a function of  $\bm{p^\text{D}}$ and $\bm{\rho}$ as $$R_{l}^\text{D}(\bm{p^\text{C}},\bm{p^\text{D}},\bm{\rho}) =\sum_{k\in\mathscr{K}}\log(1+\frac{\rho_{kl}p_{kl}^\text{D}}{{a_{kl}+b_{kl}p_{kl}^\text{D}}})\triangleq\hat{R}_{l}^\text{D}(\bm{p^\text{D}},\bm{\rho}),$$ where $a_{kl}\triangleq \sigma_N^2/h_l^\text{D}+(2^{R_{\min}^\text{C}}-1)h_{kl}^\text{CD}\sigma_N^2/h_l^\text{D}h_k^\text{CB}$ and $b_{kl}\triangleq (2^{R_{\min}^\text{C}}-1)h_{kl}^\text{CD}h_l^\text{DB}/h_l^\text{D}h_k^\text{CB}$.

From the above discussion, Problem (\ref{Mresource}) is equivalent to the following optimization problem
\begin{eqnarray*}\label{Mtraffic}
	\hspace{7.25em}\max_{\{\bm{p^\text{D}},\bm{\rho}\}} \min_{l\in \mathscr{L} } \hat{R}_{l}^\text{D}(\bm{p^\text{D}},\bm{\rho}),\hspace{7em}
	\eqref{Mtraffic_obj} \label{Mtraffic_obj}
\end{eqnarray*}
\begin{subequations}
	subject to (\ref{resource_sub1}),  (\ref{resource_sub2}),  (\ref{resource_sub3}), and
	\begin{align}
	p_{kl}^\text{D}\leq p_{kl}^{\max}, \quad{\forall}k\in \mathscr{K}, l\in \mathscr{L}.\label{traffic_sub1}
	\end{align}
\end{subequations}

\subsection{Optimal Branch-and-Bound Algorithm}
We propose to use the B\&B algorithm to solve Problem (\ref{Mtraffic}). Since the integer variable in Problem (\ref{Mtraffic}) is binary, the B\&B algorithm can attain the globally optimal solution by iteratively searching a binary tree.  Each node $n$ in the tree is associated with a nonlinear sub-problem of Problem (\ref{Mtraffic}), where the constraints of the integer variables, i.e., constraint (\ref{resource_sub1}), are modified. Specifically, some integer variables are determined while others are undetermined and relaxed into continuous variables within $[0,1]$. By solving the corresponding nonlinear problem at node $n$, its local upper bound, $b_U^n$, can be obtained, because the feasible region of the corresponding nonlinear problem is larger than that of the original MINLP problem.  The searching process of the B\&B algorithm contains three iterative steps.
\begin{itemize}
	\item Node selection: selecting a node from the unvisited node list of the tree.
	\item Evaluation: solving the corresponding nonlinear sub-problem of the node to obtain its local upper bound, $b_U^n$.
	\item Fathom decision: using the local upper bound, $b_U^n$, and the global lower bound, $b_L$, which is  the optimal value of the objective function by far, to decide whether the node should be fathomed.
\end{itemize}
The searching process comes to an end by iteratively repeating the above three steps until the node list is empty.

In the following, we will first solve the corresponding nonlinear problem at each node, and then develop the complete B\&B algorithm for Problem (\ref{Mtraffic}).

\subsubsection{Local Upper Bound Algorithm}
As mentioned before, at each node of the binary tree generated by the B\&B algorithm, some channel allocation indicators are determined while others are relaxed into a continuous variable within $[0,1]$. We use  $\bm{\rho}^d$ to denote the matrix that stores the values of the determined channel allocation indicators. Then the nonlinear sub-problem of Problem (\ref{Mtraffic}) corresponding to node $n(\bm{\rho}^d)$ can be written as
\begin{eqnarray*}\label{Ma}
	\hspace{6.5em}\max_{\{\bm{p^\text{D}},\bm{\hat{\rho}}\}} \min_{l\in \mathscr{L} } \hat{R}_{l}^\text{D}(\bm{p^\text{D}},\bm{\hat{\rho}},\bm{\rho}^d),\hspace{6.25em}
	\eqref{a_obj} \label{a_obj}
\end{eqnarray*}
\begin{subequations}
	subject to
	\begin{align}
	\hat{\rho}_{kl} \in [0,1], \quad{\forall}k\in \mathscr{K}, l\in \mathscr{L},\label{a_sub1}
	\end{align}
	\begin{align}
	\sum_{l\in \mathscr{L}}\hat{\rho}_{kl}\leq 1, \quad{\forall}k\in \mathscr{K},\label{a_sub2}
	\end{align}
	\begin{align}
	\hat{\rho}_{kl} = \bm{\rho}^d_{kl}, \quad{\forall} \{k,l\}\in \mathscr{D}_n,\label{a_sub3}
	\end{align}
	\begin{align}
	\sum_{k\in \mathscr{K}}\hat{\rho}_{kl}p_{kl}^\text{D} \leq P^\text{D}_{\max}, \quad{\forall}l\in \mathscr{L},\label{a_sub4}
	\end{align}
	\begin{align}
	p_{kl}^\text{D}\leq p_{kl}^{\max}, \quad{\forall}k\in \mathscr{K}, l\in \mathscr{L},\label{a_sub5}
	\end{align}
\end{subequations}
where $\mathscr{D}_n$ is the index set of determined channel allocation indicators. It is obvious that the solution to Problem (\ref{Ma}) serves as an upper bound of Problem (\ref{Mtraffic}).

We introduce a new variable $s_{kl} = p_{kl}^\text{D}\hat{\rho}_{kl}$. Then
\begin{align}
p_{kl}^\text{D}=\left\{\begin{array}{rcl}
0, & &{\hat{\rho}_{kl}=0,}\\
s_{kl}/\hat{\rho}_{kl},& &{\rm otherwise}. \\
\end{array} \right. \label{sp}
\end{align}
Then Problem (\ref{Ma}) can be modified into
\begin{eqnarray*}\label{Mb}
	\hspace{9em}\max_{\{\eta,\bm{s},\bm{\hat{\rho}}\}} \eta(\bm{\rho}^d),\hspace{9.2em}
	\eqref{b_obj} \label{b_obj}
\end{eqnarray*}
\begin{subequations}
	subject to (\ref{a_sub1}), (\ref{a_sub2}), (\ref{a_sub3}), and
	\begin{align}
	\sum_{k\in \mathscr{K}}s_{kl} \leq P^\text{D}_{\max}, \quad{\forall}l\in \mathscr{L},\label{b_sub4}
	\end{align}
	\begin{align}
	s_{kl}\leq \hat{\rho}_{kl} p_{kl}^{\max}, \quad{\forall}k\in \mathscr{K}, l\in \mathscr{L},\label{b_sub5}
	\end{align}
	\begin{align}
	\hat{R}_{l}^\text{D}(\bm{s},\bm{\hat{\rho}})\geq \eta, \quad{\forall}l\in \mathscr{L},\label{b_sub6}
	\end{align}
\end{subequations}
where \begin{align}
\hat{R}_{l}^\text{D}(\bm{s},\bm{\hat{\rho}}) = \sum_{k\in \mathscr{K}}\log(1+\frac{\hat{\rho}_{kl}s_{kl}}{a_{kl}\hat{\rho}_{kl}+b_{kl}s_{kl}}).
 \label{concave}
\end{align}

Problem (\ref{Mb}) is obviously equivalent to Problem (\ref{Ma}). Furthermore, we can prove that (\ref{concave}) is a concave function \cite{convex} and Problem (\ref{Mb}) is a convex optimization problem. Therefore, the interior-point method  can be used to solve it and thus the local upper bound of each node can be obtained, which will be used to decide whether a node should be fathomed during the B\&B searching process.

\subsubsection{B\&B Algorithm}
After obtaining the local upper bound of each node, we can develop the B\&B algorithm for Problem (\ref{Mtraffic}), which is summarized in Table I. For simplicity,  we adopt the depth-first-search (DFS) \cite{DFS} as the node selection rule and always choose the first undetermined channel indicator for variable selection process. However, the algorithm in Table I is impractical  due to its exponential computational complexity, i.e., $O(2^{KL})$, which motivates us to accelerate it by ML techniques.
\begin{table}[h]
\setlength{\abovecaptionskip}{-2pt}
\setlength{\belowcaptionskip}{-6pt}
\caption{Optimal B\&B Algorithm}
\begin{algorithm}[H]
\caption{Optimal B\&B Algorithm}
\label{A1}
{\small
	\begin{algorithmic}[1]
	\State \textbf{initialization}
	\begin{itemize}
		\item Set node list: $\mathscr{N}\leftarrow\{n([ ])\}$.
		\item Set global lower bound: $b_L = -\infty$.
		\item Set optimal solution: $\bm{\rho^*}\leftarrow  None$, $\bm{{p^\text{D}}^*}\leftarrow  None$.
	\end{itemize}
	\While{$\mathscr{N} \neq \varnothing$}
	\State \textbf{Node Selection:} pop the first node $n(\bm{\rho}^d)$ in $\mathscr{N}$.
	\State \textbf{Evaluation:} use interior-point method solve Problem (\ref{Mb}) \indent   corresponding to $n(\bm{\rho}^d)$.
	\State \textbf{Fathom Decision:}
	\If {Problem (\ref{Mb}) corresponding to $n(\bm{\rho}^d)$ is infeasible}
	\State Go to step 2.
	\Else
	\State Get $\eta,\bm{\hat{\rho},p^\text{D}}$. Set $b_U^n = \eta$.
	\If{$\bm{\hat{\rho}}$ is integral}
	\State An integer solution is found.
	\If{$b_U^n>b_L$}
	\State Update global lower bound and optimal solution.
	\State $b_L = b_U^n$,  $\bm{\rho^*}=\bm{\hat{\rho}}$, $\bm{{p^\text{D}}^*}=\bm{p^\text{D}}$.
	\EndIf
	\State Go to step 2.
	\Else
	\If{$b_U^n<b_L$}
	\State Prune node $n(\bm{\rho}^d)$.
	\State Go to step 2.
	\Else
	\State Branch on node $n(\bm{\rho}^d)$.
	\State \textbf{Variable Selection:} branch on the first  undeter
	\indent \indent\indent \indent -mined variable in $\bm{\hat{\rho}}$.
	\State Add the new nodes into $\mathscr{N}$.
	\State Go to step 2.
	\EndIf
	\EndIf
	\EndIf
	\EndWhile
    \end{algorithmic}}
\end{algorithm}
\vspace{-3em}
\end{table}

%%% ------------------------- \input{sub/conclusion.tex} --------------
\section{Accelerating by Imitation Learning}
In this section, we will use the ML technique, specifically the imitation learning method, to accelerate the algorithm in Table I. We will first discuss the optimal auxiliary prune policy and then give a  brief introduction to the concept of imitation learning. Finally, we will introduce how to use it in our specific problem and illustrate the detailed training process.

\subsection{Optimal Auxiliary Prune Policy}
During the B\&B searching process  in Table I, there are two main goals: finding an optimal solution and guaranteeing its optimality by searching all feasible solutions and comparing with it. Most of the time is consumed in the latter one and a good prune policy can significantly reduce the computational complexity. The more nodes are pruned, the less time is consumed.

The original prune policy of the algorithm in Table I only includes three cases:
\begin{itemize}
	\item The sub-problem is infeasible as shown in line 6 of Table I. If the relaxed nonlinear sub-problem corresponding to $n(\bm{\rho}^d)$ is infeasible, the related MINLP problem is also infeasible and then the node $n(\bm{\rho}^d)$ is fathomed.
	\item A feasible solution is found as shown in line 10 of Table I. If $\bm{\hat{\rho}}$ is an integral vector, the result is also a feasible solution of the related MINLP problem and then the node $n(\bm{\rho}^d)$ is fathomed.
	\item The local upper bound is smaller than the current global lower bound as shown in line 18 of Table I. If $\bm{\hat{\rho}}$ is not an integral vector and $b_U^n<b_L$, then the node $n(\bm{\rho}^d)$ would not lead to a better solution and is fathomed.
\end{itemize}

Besides the three cases mentioned above, all the non-optimal nodes should be pruned to accelerate the B\&B algorithm. We can achieve this goal by using the method in \cite{Hhe,LORA}, where  the aforementioned original prune policy in the B\&B algorithm has been replaced with a learned policy that would prune all the non-optimal nodes. However, it is redundant to learn to prune the fathomed nodes since the original prune policy is compact and explicit. Therefore, we propose to simplify the learning task by keeping the original prune policy and learning an auxiliary prune policy to reinforce it.  The auxiliary prune policy is supposed to prune the non-optimal nodes that are not fathomed instead of all the non-optimal ones. Our proposed simplification about learning task can reduce the complexity of target model. Generally, simpler model has smaller Vapnik-Chervonenkis (VC) dimension, which leads to fewer needed training samples and speeds up the training process consequently. In this way, the best auxiliary prune policy, $\pi_p^*$, should discard all non-optimal nodes that have not been fathomed. In the following, we introduce how to use imitation learning to learn the best auxiliary prune policy, $\pi_p^*$.

\subsection{Basic Imitation Learning}
Imitation learning has been widely used in sequential decision problems, where the learner tries to mimic an expert's action, i.e., an oracle, to achieve the best performance in a supervised manner \cite{imilearn}. Generally, a sequential decision problem is defined by a state space, $\mathscr{S}$, an action space, $\mathscr{A}$, and a policy space, $\Pi$. A policy $\pi \in \Pi$ reflects the mapping relations between the states and actions, i.e., $\pi(s) = a$. The oracle, $\pi^*$, is a special policy that provides the optimal action, $a^*$, for any possible state $s\in \mathscr{S}$. The goal of imitation learning is to find a policy that mimics the oracle's actions.

We shall note that imitation learning is different from reinforcement learning. Although the goal of reinforcement learning is also to learn a policy, the best policy for reinforcement learning is unknown and should be learned  by interacting with the environment in an unsupervised manner. Whereas, the best policy in imitation learning is known and can be learned in a supervised manner. Imitation learning better fits accelerating the B\&B algorithm than reinforcement learning.

The B\&B searching process in Table I can be formulated as a sequential decision-making process. At each node encountered during the searching process, we need to decide whether it is pruned using the prune policy and the decision can influence the whole B\&B searching process. Here, the state space, $\mathscr{S}$, is the set of all the visited nodes with the corresponding global lower bound and the current optimal solution. The action space, $\mathscr{A}$, is $\{prune, branch\}$ and the policy space is defined as $\Pi_p$.

Let $\phi: \mathscr{S} \rightarrow \mathbb{R}^q$ be the feature mapping that finds a $q$-dimension feature-vector description for any state $s\in \mathscr{S}$. Given that our goal is to learn the best auxiliary prune policy, $\pi_p^*$, i.e., learn the  optimal action, $a^*$, for any possible state $s\in \mathscr{S}$,  the imitation learning problem can be converted into a supervised learning problem: the policy takes a feature-vector description of the state $s$ as input and tries to predict the oracle action, $a^*$, as output. Moreover, because the action space, $\mathscr{A}$, is only two-dimensional, the imitation learning problem can be further converted into a binary classification problem.

\subsection{Imitation Learning for Auxiliary Prune Policy}
Now, we utilize the imitation learning method to learn the optimal auxiliary prune policy.  As mentioned above, the imitation learning problem can be converted into a binary classification task with appropriate  feature mapping, $\phi$. In this part, we will first discuss how to design appropriate feature mapping, and then introduce how to use SVM to train the binary classifier.
\subsubsection{Feature  Design} Feature design is very important for training classifiers. We need to dig out the features that are closely related to state $s\in \mathscr{S}$. In \cite{Hhe}, problem-independent features have been used for general MILP problems, which can be extended to the MINLP problems. On the other hand, problem-dependent features have been exploited in\cite{LORA} in addition to the problem-independent ones to further improve the performance of the classifier. Here, we adopt both kinds of features.

\textit{\romannumeral1. Problem-Independent Features:}

Problem-independent features focus on the structure of the binary tree generated by the B\&B algorithm, which can be used in all MINLP problems regardless of applications. It consists of the following three categories \cite{Hhe}.
\begin{itemize}
	\item Node features: State $s$ is highly relevant to current visiting node $n$, thus features computed from current node $n$ are essential. They include the depth of node $n$, the plunge depth of node $n$, and the local upper bound, $b_U^n$, of node $n$.
	\item Branching feature: State $s$ also depends on the branching variable leading to current node $n$, which is chosen at the variable selection step of the parent of node $n$. It is less important than node features in the auxiliary prune policy learning problem \cite{Hhe}. Here we use the value of the branching variable as the only branching feature.
	\item Tree features: Features computed from the binary tree are also important for describing state $s$, which includes current global lower bound, $b_L$, and the number of  solutions obtained so far.
\end{itemize}

Most of the features mentioned above vary largely for different problems. For example,  the objective values of problems with different CSI vary a lot. On the other hand, the plunge depths of nodes vary with problem sizes. Therefore, we normalize the features involving bound with the local upper bound at the root node as well as those involving depth with the maximum depth of the binary tree. This normalization process makes the above features size-independent, i.e., independent on the numbers of CUs and D2D pairs, which can help improve the generalization ability of the proposed learned prune policy.

\textit{\romannumeral2. Problem-Dependent Features:}

Problem-dependent features, as the name implies, are closely related to the specific problem we solve. To choose appropriate problem-dependent features, we need to check the original Problem (\ref{Mresource}) and find the key parameters related to the channel and power allocation for D2D communications. It is obvious that CSI and power constraints are two key factors. Therefore, we try to design problem-dependent features according to these two factors in the following.
\begin{itemize}
	\item CSI feature: At node $n$, there are four different kinds of CSI involved, i.e., $h_{kl}^\text{CD}$, $h_{k}^\text{CB}$, $h_{l}^\text{D}$, and $h_{l}^\text{DB}$, when the first undetermined channel allocation indicator is $\hat{\rho}_{kl}$. In Proposition 2, we have introduced two new variables $a_{kl}$ and $b_{kl}$ that combine these four kinds of CSI with the data rate of D2D pair $l$ on the channel of CU $k$, i.e., $R_{kl}^\text{D}$. According to (\ref{abr}), we can formulate the CSI feature as $\log(1+1/(a_{kl}+b_{kl}))$. On the other hand, both $a_{kl}$ and $b_{kl}$ are related to $R_\text{min}^C$ that varies across different problems. Therefore, we normalize the CSI feature with  $R_\text{min}^C$  and get the size-independent CSI feature as $f(CSI)=(\log(1+1/(a_{kl}+b_{kl})))/R_\text{min}^C$.
	\item Power feature: At node $n$, when the first undetermined channel allocation indicator is $\hat{\rho}_{kl}$, we formulate a function of $p_{kl}$ as the power feature. According to (\ref{traffic_sub1}), we can denote the power feature as $p_{kl}^\text{max}$. Because $p_{kl}^\text{max}$ also varies across different problems, the normalized power feature can be similarly formulated as
	$g(p_{kl})= KLp_{kl}^\text{max}/\sum\limits_{k=1}^K \sum \limits_{l=1}^Lp_{kl}^\text{max}$.
\end{itemize}

These two aforementioned problem-dependent features are also size-independent.  Combined with the size-independent problem-independent features, we can expect that the auxiliary prune policy we learn from the oracle of specific scenarios can be applied to scenarios with different sizes, i.e., different numbers of CUs and D2D pairs, which will be confirmed by the test results in Section V.

\subsubsection{Binary Classifier Learning}
After choosing an appropriate feature mapping, $\phi$, we can handle the binary classification problem. We use SVM to train the classifier, which is widely used for classification. The SVM aims to find  a mapping of training examples so that they can be divided by a clear gap that is as wide as possible in the new space \cite{svm}. We stack the  eight features mentioned above in an 8-dimension vector as the input and use the corresponding action, $a^*$, as the output. Moreover, we denote \emph{branch}, i.e., optimal node, as 1, and \emph{prune}, i.e., non-optimal node, as 0.

During the B\&B searching process, mistakes made at early stages are more serious. For example, the root node is always supposed to be classified as branching. If it is misclassified as pruning, the searching process ends and no feasible solution can be found. On the other hand, it is more serious to prune the optimal nodes by mistake than to keep non-optimal nodes. If an optimal node is misclassified as non-optimal and is pruned, the optimal solution cannot be found either. However, if a non-optimal node is misclassified as optimal, just more time is consumed but the optimality of result is still preserved.

To deal with the first problem mentioned above, we place higher weights on the training examples from the nodes with small depth. Because the subtree size decreases exponentially with the node's depth, the weight related to node's depth, $\omega_1$, is also supposed to decrease exponentially for nodes at different depths. Therefore, we set $\omega_1$ for a node with depth $d$ as $\omega_1=Ae^{\frac{-Bd}{D}}$, where $D$ is the maximum depth of the B\&B tree, $A$ and $B$ are the parameters to be tuned during the training process. Specifically, we set $A=5$ and $B=2.68$ according to \cite{Hhe}, which enables the weight of the root node to be $5$ and the weights for nodes at different depths decrease at the rate $2.68/D$. Nevertheless, different values of $A$ and $B$ will impose different impacts on the results. Increasing $A$ will reinforce the optimality by placing higher weights while increasing $B$ will increase the decay rate and differentiate more clearly for nodes at different depths. In practice, we can choose appropriate $A$ and $B$ according to different goals.

As for the second problem mentioned above, we  place higher weights on the training examples from optimal nodes and denote the weight related to node's optimality as $\omega_2$. We set $\omega_2$ as $1$ for non-optimal nodes and tune it for optimal ones from $\{1,2,4,8\}$ during the training process. There are two main benefits of placing $\omega_2$:  avoiding the serious optimal node misclassification issue mentioned above and dealing with the dataset imbalance issue. Dataset imbalance issue means the number of the optimal nodes is much smaller than that of the non-optimal ones, which can be mitigated by placing higher weights at the optimal nodes. Combined these two parts mentioned above, the total weight, $\omega$,  for each training sample is the product of $\omega_1$ and $\omega_2$, i.e., $\omega =\omega_1\times \omega_2$.

\subsection{Iterative Training Process}
The binary classifier is trained in the supervised manner. The general method puts all the nodes searched by the algorithm in Table I into the training set. However, this method is memory-consuming because the number of nodes increases exponentially. Therefore, we adopt an iterative
training algorithm, dataset aggregation (DAgger) \cite{dagger,Hhe}, to achieve more efficient training.

DAgger is widely used in the imitation learning to provide a learning reduction with
strong performance guarantee \cite{dagger}. It proceeds by collecting a dataset
at each iteration using current policy and then trains a new policy with the aggregation of all collected datasets. The iterative dataset collection process can enhance the generalization ability of the model to some extent.

The detailed training procedures for imitation learning using DAgger are shown in Tables II and III. At the first iteration, we use the oracle, $\pi_p^*$, as the initial policy and denoted as $\pi_p^{(1)}$. We search each problem in the problem set, $\mathscr{Q}$, with  $\pi_p^{(1)}$ and collect data into the training dataset, $\mathscr{T}$. Then we use  $\mathscr{T}$ to train SVM and get a new learned policy denoted as  $\pi_p^{(2)}$. We repeat this process for $M$ rounds and choose the policy, $\pi_p^{(m)}$, that performs the best on the validation set.

Note that, some steps in Algorithm 3 are similar with those in Algorithm 1 because the B\&B tree needs to be searched in both algorithms. However, Algorithm 3 is mainly developed to collect the training data for the iterative training process in Table II. The new learned policy has been adopted in the Algorithm 3 for pruning decision, which makes it different from that in Algorithm 1.

\begin{table}[h]
	\setlength{\abovecaptionskip}{-2pt}
	\setlength{\belowcaptionskip}{-6pt}
	\caption{Training Process for Imitation Learning with DAgger}
	\label{A2}
	\begin{algorithm}[H]
		\caption{Training Process for Imitation Learning with DAgger}
		{\small
	\begin{algorithmic}[1]
		\State \textbf{initialization}
		\begin{itemize}
			\item Set policy: $\pi_p^{(1)}=\pi_p^*$.
			\item Set training dataset: $\mathscr{T}=\emptyset$.
		\end{itemize}
		\For {$m = 1$ \textbf{to} $M$}
		\For {problem $Q$ \textbf{in}  $\mathscr{Q}$}
		\State $\mathscr{T}^{(Q)} \leftarrow$ DataCollection($Q, \pi_p^{(m)}$)
		\State $\mathscr{T} \leftarrow \mathscr{T} \cup \mathscr{T}^{(Q)} $
		\EndFor
		\State $\pi_p^{(m+1)} \leftarrow$ train SVM (classifier) using $\mathscr{T}$.
		\EndFor
		\State \textbf{return} Best $\pi_p^{(m)}$ on the validation set.
    \end{algorithmic}}
\end{algorithm}
\vspace{-1em}
\end{table}

Here, we make some improvements on the training process to accelerate it and also deal with the dataset imbalance problem. Specifically, we only collect the nodes that are visited under current policy and are not fathomed by the original B\&B prune policy during the data collection process shown in Table III. It is different from the  data collection process in \cite{Hhe} and \cite{LORA}, which includes all the visiting nodes into the dataset. There exist two main reasons to make this improvement.
\begin{itemize}
	\item  When visiting nodes that are fathomed by the original B\&B prune policy, the learned auxiliary prune policy is not needed. Therefore, ignoring those fathomed nodes would not induce loss in performance but can speed up the training process of imitation learning. Note that if we discard the original prune policy and learn a new one to replace it as proposed in \cite{Hhe,LORA}, the fathomed nodes cannot be neglected during the data collection process.  Therefore,  the improvement on the training process is directly led by keeping the original prune policy.
	\item The nodes that are fathomed by the original B\&B prune policy are non-optimal. Therefore, discarding these nodes from the dataset can decrease the number of non-optimal nodes  and mitigate the dataset imbalance issue mentioned above to some extent.
\end{itemize}

\begin{table}[h]
	\setlength{\abovecaptionskip}{-2pt}
	\setlength{\belowcaptionskip}{-6pt}
	\caption{Data Collection Algorithm: DataCollection($Q, \pi_p$)}
	\label{A3}
	\begin{algorithm}[H]
		\caption{DataCollection($Q, \pi_p$)}
		{\small
			\begin{algorithmic}[1]
				\State \textbf{initialization}
				\begin{itemize}
					\item Set node list: $\mathscr{N}_Q\leftarrow\{n([ ])\}$.
					\item Set dataset: $\mathscr{D} = \emptyset$.
				\end{itemize}
				\While{$\mathscr{N}_Q \neq \varnothing$}
				\State \textbf{Node Selection:} pop the first node $n_Q(\bm{\rho}^d)$ in $\mathscr{N}_Q$.
				\State \textbf{Evaluation:} use interior-point method solve Problem (\ref{Mb})  \indent corresponding to $n_Q(\bm{\rho}^d)$.
				\State \textbf{Fathom Decision:}
				\If {Node $n_Q(\bm{\rho}^d)$ is fathomed by original B\&B}
				\State Go to step 2.
				\Else
				\State  $\mathscr{D} \leftarrow \{\phi(n_Q(\bm{\rho}^d)), \pi_p^*(n_Q(\bm{\rho}^d))\}$
				\If {$\pi_p(n_Q(\bm{\rho}^d)) = branch$}
				\State Branch on node $n_Q(\bm{\rho}^d)$.
				\State \textbf{Variable Selection:} branch on the first  undetermined \indent \indent\indent variable in $\bm{\hat{\rho}}$.
				\State Add the new nodes into $\mathscr{N}_Q$.
				\EndIf
				\State Go to step 2.
				\EndIf
				\EndWhile
				\State \textbf{return} $\mathscr{D}$
		\end{algorithmic}}
	\end{algorithm}
\vspace{-2em}
\end{table}

%%% ------------------------- \input{sub/conclusion.tex} --------------
\section{Performance Test Results}
In this section, we will test the performance of the proposed imitation learning method on accelerating the algorithm in Table I. All the codes are implemented in python 3.6  except the interior-point algorithm that is implemented in Matlab.  They are implemented on a computer station with one 4-core Intel processor and 128 GB of memory. To avoid over-fitting, we set $M=4$ for all the imitation learning training process and use LIBSVM \cite{libsvm} for the step of training  SVM (classifier) in the following. Given that the main goal of this paper is using a ML technique to accelerate the B\&B algorithm, we mainly test the performance of the ML technique with specific focus on scalability and generalizability. Moreover, to honor the tradition in the ML community, all the testing results are presented in tables with specific numerical values.

\subsection{System Setup}
We consider a single-cell network with a radius of 500 m as in Fig. \ref{fig:sysmodel}. The eNB is located in the center of the cell and the CUs are distributed uniformly in the cell. According to \cite{yuwei},  the transmitter of each D2D pair is also distributed uniformly in the cell and the corresponding receiver is distributed in a disk centered by the transmitter with fixed uniform link distance distribution between $r_{\min}$ and $r_{\max}$. Our simulation parameters are summarized in Table \ref{table1}.

\begin{table}[ht]
	\small
	\caption{Simulation Parameters}
	\label{table1}
	\centering
	\begin{tabular}{|c|c|}
		\hline
		Parameter & Value  \\
		\hline
		\hline
		Cell radius & 500 m \\
		\hline
		D2D distance, $r_{\min}$, $r_{\max}$ & 15 m, 50 m \\
		\hline
		Noise spectral density & -174 dBm/Hz\\
		\hline
		\tabincell{c}{Path loss model \\ for cellular links} & 128.1+37.6log(d[km])\\
		\hline
		\tabincell{c}{Path loss model \\for cellular D2D links} & 148+40log(d[km])\\
		\hline
		Shadowing standard deviation & 10 dB \\
		\hline
		\tabincell{c}{Maximum transmitter \\ power of CU, $P^\text{C}_{\max}$} & 20 dBm \\
		\hline
		\tabincell{c}{Maximum transmitter \\power of D2D pair, $P^\text{D}_{\max}$} & 20 dBm \\
		\hline
		Minimum data rate of CU, $R^\text{C}_{\min}$ & 2 bit/s/Hz\\
		\hline
	\end{tabular}
\end{table}

\subsection{Influence of the Number of Training Samples}
An important test of the proposed method is how many training samples are enough to learn a good auxiliary prune policy since training samples are always difficult and expensive to get, especially in wireless networks where a large number of real data are hard to obtain.

We test the performance of the proposed method with different numbers of training samples for scenarios with different numbers of CUs and D2D pairs. Because we usually have a small labeled dataset for wireless problems, we set the ratio between the number of training samples and that of testing samples as 10:1 for the subsequent test and use 20 testing samples for each scenario.  During the testing stage, the learned model will be applied to every encountered node while searching the B\&B tree for each testing sample. Given that the number of nodes of the B\&B tree is exponentially increasing, 20 testing samples include hundreds even thousands of testing nodes, which are sufficient for performance evaluation.

The testing results are shown in Table \ref{table2}. We use four metrics to measure the performance of the proposed method. \emph{Ogap} is the optimality gap, which means the gap between the optimal objective function and the one achieved by the proposed imitation learning method.  \emph{Speed} refers to the speedup with respect to the original B\&B searching process, which is computed as the ratio between the number of nodes explored without and with the learned auxiliary policy.  Note that all the running time is tested with the same hardware set. \emph{Optimal recognition rate} is the accuracy rate of the learned policy on recognizing optimal nodes, which is positively related to \emph{ogap}. \emph{Extra prune rate} is the percentage of the non-optimal nodes that are not fathomed by the original B\&B prune policy but can be recognized by the learned auxiliary policy. It is obvious that \emph{speed} and \emph{extra prune rate} are also positively correlated. Note that the values of \emph{ogap} and \emph{speed} are the average of the 20 testing samples while those of \emph{optimal recognition policy} and \emph{extra prune policy} are the average of the hundreds and even thousands of testing nodes of the B\&B trees for  the 20 testing samples.
\begin{table}
	\small
	\caption{Performance of Imitation Learning with Different Numbers of Training Samples}
	\label{table2}
	\vspace{-1em}
	\centering
	\subtable[$K=5$, $L=2$]{
		\begin{tabular}{|c|c|c|c|c|}
			\hline
			\tabincell{c}{Number of \\training samples} & 50 & 100 & 150 & 200 \\
			\hline
			Ogap  & 3.88\%& 3.23\%& 2.27\% & 2.01\%\\
			\hline
			Speed& 2.50x& 2.21x& 2.17x& 2.06x\\
			\hline
			\tabincell{c}{Optimal recognition\\ rate}  & 89.22\%& 93.27\%& 93.33\%& 93.38\%\\
			\hline
			Extra prune rate & 30.88\% & 30.32\%& 29.00\%& 28.20\%\\
			\hline
		\end{tabular}
		\label{tab2:firsttable}
	}
	\qquad
	\subtable[$K=7$, $L=2$]{
		\begin{tabular}{|c|c|c|c|c|}
			\hline
			\tabincell{c}{Number of \\training samples}  & 50 & 100 & 150 & 200 \\
			\hline
			Ogap& 7.81\%&7.62\%& 7.06\% & 6.96\% \\
			\hline
			Speed& 9.51x& 4.91x& 4.89x& 4.82x\\
			\hline
			\tabincell{c}{Optimal recognition\\ rate} & 80.46\%& 86.92\%& 88.65\%& 90.21\%\\
			\hline
			Extra prune rate  & 18.07\%& 38.24\%& 41.11\%& 40.96\% \\
			\hline
		\end{tabular}
		\label{tab2:secondtable}
	}
	\qquad
\subtable[$K=5$, $L=3$]{
	\begin{tabular}{|c|c|c|c|c|}
		\hline
		\tabincell{c}{Number of \\training samples}  & 50 & 100 & 150 & 200 \\
		\hline
		Ogap& 14.54\%&13.08\%& 12.54\% & 11.92\% \\
		\hline
		Speed& 15.79x& 15.45x& 13.91x& 11.66x\\
		\hline
		\tabincell{c}{Optimal recognition\\ rate} & 83.33\%& 85.46\%& 86.21\%&86.89\%\\
		\hline
		Extra prune rate  & 42.22\%& 48.47\%& 50.00\%& 50.91\% \\
		\hline
	\end{tabular}
	\label{tab3:thirdtable}}
\end{table}

Our algorithm is expected to attain a small objective function gap while reducing the computational complexity as much as possible. As shown in Table \ref{table2}, the proposed method can speed up the B\&B searching process by 2.06 times with only 2.01\% loss of accuracy for the scenario with $K=5$, $L=2$, and can speed up the B\&B searching process by 4.82 times with 6.96\% loss of accuracy for the scenario with $K=7$, $L=2$, both with only 200 training samples. For the more complicated scenario with $K=5$, $L=3$, the proposed method can speed up the B\&B searching process by 11.66 times with 11.92\% loss of accuracy while using 200 training samples. If reducing to 50 training samples, the proposed method can still achieve low performance gaps of 3.88\%, 7.81\% and 14.54\% for the three scenarios, respectively. Comparing to the one million training samples used in \cite{shi}, the proposed imitation learning method can achieve a good performance with a small set of training samples, and thus is more feasible for wireless networks. On the other hand, \emph{ogap} and \emph{speed} both decrease with the increase of the number of training samples. As we can imagine, using more training samples will increase the optimality but will also induce more computational complexity, which indicates the optimality and computational complexity trade-off issue. Small training sets lead to a policy with lower computational complexity but also lower optimality, while larger training sets lead to the opposite results. This can  help us choose appropriate numbers of training samples in practice according to our specific goals instead of using as much training samples as possible.  Furthermore, with the increase on the number of training samples, the performance for the scenario with $K=5$, $L=2$ improves very slowly while those for the scenarios with $K=7$, $L=2$ and $K=5$, $L=3$ improve at a stable speed. And the achieved optimality for the scenario with $K=5$, $L=2$ is always higher than those for the scenarios with $K=7$, $L=2$ and $K=5$, $L=3$ when using the same number of training samples. These results indicate that the policy for the scenario with $K=5$, $L=2$ is fully trained but those for the scenarios with $K=7$, $L=2$ and $K=5$, $L=3$ are not yet. Larger training sets are needed for larger scale problems than for smaller scale problems to achieve the same optimality.

\subsection{Influence of Feature Selection}
Feature selection is very important for training classifier and we have used eight problem-independent and problem-dependent features. As mentioned in Section IV-C, problem-independent features are general for all MINLP problems while problem-dependent ones need to be carefully selected according to specific applications. We will test the importance of both kinds of features in the following to obtain some insights about feature selection in practice.

\begin{table}
	\small
	\caption{Performance of Imitation Learning with Different Sets of Features for the Scenario with $K=5$, $L=2$}
	\label{table3}
	\centering
	\begin{tabular}{|c|c|c|}
		\hline
		Feature set& \tabincell{c}{Problem-independent\\features only} & \tabincell{c}{Two kinds of \\features combined} \\
		\hline
		Ogap &3.11\% &2.01\% \\
		\hline
		Speed &1.93x  & 2.06x\\
		\hline
		\tabincell{c}{Optimal recognition\\ rate}  &91.45\% &93.38\% \\
		\hline
		Extra prune rate  &26.48\% &28.20\%\\
		\hline
	\end{tabular}
\end{table}
We test on the scenario with $K=5$, $L=2$ using 200 training samples and 20 testing samples. The results of training with problem-independent features only and with both kinds of features are summarized in Table \ref{table3}. From the table, the proposed method can speed up the B\&B searching process by 1.93 times with only 3.11\% performance loss with only problem-independent features while adding the problem-dependent features can improve the optimality and reduce computational complexity simultaneously. This result suggests the necessity of the problem-dependent features in reinforcing the performance.

Furthermore, we rank all the eight features using \emph{F-test}, i.e., joint hypotheses test, which is widely applied in feature selection for ML. We find that the most important two features are the local upper bound, $b_U^n$, and the global lower bound, $b_L$. The learned policy is inclined to branch on the nodes whose difference between $b_U^n$ and $b_L$ are very small, which means these nodes have very large probabilities to be the optimal ones. The less two important features are two problem-dependent features, which coincides with the fact mentioned above that the proposed method can also achieve a satisfactory performance without  problem-dependent features.

Inspired from the above observation, we can implement the imitation learning method in a two-step paradigm in practice. First, we only use problem-independent features for training. If the optimality and the computational complexity of the learned policy is acceptable, then we can skip the intricate design process of problem-dependent features. Otherwise, we then add problem-dependent features and train a new learned policy by using both kinds of features.

\subsection{Generalization to Scenarios with Different Sizes}
Generalization ability is another important property of ML techniques. As mentioned in Section IV-C, we expect our method can be generalized to scenarios with different problem sizes. In the following, we first test whether the proposed imitation learning method has generalization ability, which includes generalization abilities to both the larger and the smaller scenarios. Then we test how strong its generalization ability is.

\begin{table}
	\scriptsize
	\caption{Cross Generalization Test on the Performance of Imitation Learning for Scenarios  with $K=5$, $L=2$ and $K=7$, $L=2$}
	\label{table4}
	\centering
	\begin{tabular}{|c|c|c|c|c|}
		\hline
		\tabincell{c}{Cross \\ generalization\\ test combination} & \tabincell{c}{Policy(5,2) \\ on \\ Problem(5,2)} & \tabincell{c}{Policy(5,2) \\ on \\ Problem(7,2)} &\tabincell{c}{Policy(7,2) \\ on \\ Problem(5,2)} & \tabincell{c}{Policy(7,2) \\ on \\ Problem(7,2)}\\
		\hline
		Ogap  &2.01\% &3.70\% &9.44\% &6.96\%\\
		\hline
		Speed &2.06x& 3.46x&3.25x & 4.82x\\
		\hline
		\tabincell{c}{Optimal\\ recognition\\ rate} &93.38\%  &90.65\%&85.29\% &90.21\%\\
		\hline
		\tabincell{c}{Extra\\ prune rate} &28.20\% &24.77\%& 47.35\% &40.96\%\\
		\hline
	\end{tabular}
\end{table}

First, we do cross generalization tests on scenarios with $K=5$, $L=2$ and $K=7$, $L=2$. We use 200 training samples and 20 testing samples for both scenarios. The results are summarized in Table \ref{table4}, where \emph{Policy($k,l$)} means the policy learned from  the scenario with $K=k$, $L=l$, and \emph{Problem($k,l$)} means the problem on the scenario with $K=k$, $L=l$. It is obvious that the proposed method has the generalization ability to the larger and the smaller scale problems. Specifically, the optimality by using \emph{Policy($5,2$)} on  \emph{Problem($7,2$)} is even higher than using \emph{Policy($7,2$)} on  \emph{Problem($7,2$)}. This result seems not intuitive but is reasonable. As mentioned above,  \emph{Policy($7,2$)} is not fully trained and more training samples are needed to improve its performance. Meanwhile, \emph{Policy($5,2$)} is fully trained and has a strong generalization ability. Therefore, \emph{Policy($5,2$)} can achieve better performance than \emph{Policy($7,2$)}  on \emph{Problem($7,2$)}. This result suggests that if there are not enough training samples for large scenario, using a fully trained policy from a small scenario can achieve better performance than training a policy with insufficient samples from the large scenario.

After checking the existence of the generalization ability of the proposed  method, we now turn to test how strong its generalization ability is, especially the generalization ability to the larger scenarios.  Given that \emph{Policy($5,2$)} is fully trained and will improve very little with more training samples, we regard \emph{Policy($5,2$)} learned with 200 training samples as the best policy. We test the performance of this policy  on scenarios with different problem sizes. For each scenario, we use 20 testing samples. The result is shown in Table \ref{table5}, where problem size \emph{($k,l$)} denotes the scenario with  $K=k$, $L=l$.  Note that, the largest scenario we consider here includes 10 CUs and 2 D2D pairs, because the training samples for larger-scale scenarios are extremely hard to obtain due to the exponentially increasing complexity of the B\&B algorithm.

From Table \ref{table5}, the performance of \emph{Policy($5,2$)} still remains acceptable even for \emph{Problem($10,2$)} whose complexity is $2^{10}$ larger than that of \emph{Problem($5,2$)}.  The above result shows that we can use the policy learned from the smaller scale problems to the larger scale problems with good performance. On the other hand, the \emph{optimal recognition rate} decreases with the increase of the problem scale. This result implies that the generalization ability is constrained by the problem size of training samples and deteriorates with the size of objective problems. This phenomenon inspires us to use mixed training strategy and soft-decision algorithm to further improve the training performance in the next section.

\begin{table}
	\small
	\caption{Performance of Imitation Learning for Scenarios  \protect\\ with Different Sizes by Using Policy(5,2)}
	\label{table5}
	\centering
	\begin{tabular}{|c|c|c|c|c|c|c|c|}
		\hline
		Problem size & (7,2) & (5,3) & (8,2) & (10,2)\\
		\hline
		Ogap & 3.70\%& 11.53\%& 5.46\% & 7.69\%\\
		\hline
		Speed & 3.46x&  4.87x&  7.25x&   22.13x\\
		\hline
		\tabincell{c}{Optimal recognition \\rate} & 90.65\%& 90.47\%&  89.78\%&   84.91\%\\
		\hline
		Extra prune rate& 24.77\%& 31.36\%& 26.87\% & 28.42\%\\
		\hline
	\end{tabular}
\end{table}

%%% ------------------------- \input{sub/conclusion.tex} --------------
\section{Discussion and Further Improvement}
Inspired by the test results in Section V, we come up with two important questions: how to strengthen the generalization ability of the proposed imitation learning method, and how to dynamically control the optimality and computational complexity trade-off. In this section, we will focus on these two questions and come up with some methods to solve them.
\subsection{Mixed Training Strategy}
As shown in Section V-D, the generalization ability is good but also limited. Policy learned from small-scale problems can be generalized to large-scale problems, but the performance deteriorates with the increase of the problem scale. It is obvious that the generalization ability is constrained by the problem scale of training samples.

In \cite{LORA}, the transfer learning method has been used to solve this problem. Its basic idea is adding few unlabeled training samples from large-scale problems and making use of an exploration policy to enhance the generalization ability. However, the exploration policy,  which is crucial for transfer learning, is always hard to design for the MINLP problems in wireless networks. Furthermore, the transfer learning method would induce higher computational complexity for training process. Therefore, we propose to use  the mixed training strategy to overcome the aforementioned shortcomings.

Specifically, if we want to solve \emph{Problem($k,l$)}, we use a training set consists of many samples from the  problems of smaller size than \emph{Problem($k,l$)} and very few samples from \emph{Problem($k,l$)}. Different from \cite{LORA}, samples from \emph{Problem($k,l$)} are also labeled.  This will not cost much overhead because it is not difficult to get very few labeled training samples from the large-scale problems in practice.  For instance, we can make use of system history record to get some optimal allocation results and use them as training samples. Since the proposed algorithm needs very few training samples, the cost of gathering historical data would be fairly low. In this way, we reinforce the generalization ability without designing complicated exploration policy and consuming more training time. The only overhead is getting few labeled training samples from the large-scale problems, which is acceptable.

We implement the mixed training strategy for each scenario with the sizes in Table \ref{table5} to test the effectiveness of the proposed strategy. For each scenario, we use a mixed training set consisting of 190 samples from the scenario with $K=5$, $L=2$ and 10 samples from itself. Moreover, we use 20 testing samples for each scenario. The results are summarized in Table \ref{table6}. Comparing Tables \ref{table5} and \ref{table6}, \emph{ogap} achieved by the  mixed training strategy is  always lower than that achieved by using the original training strategy. For some specific scenarios, it can even achieve lower computational complexity at the same time. The results suggest that the mixed training strategy achieves better optimality than using small-scale problems' training samples only. This method can achieve a trade-off between the learned policy's performance and the cost of getting training samples from the large-scale problems.

\begin{table}
	\small
	\caption{Performance of Imitation Learning for Scenarios  \protect\\ with Different Sizes by Using Mixed Training Strategy}
	\label{table6}
	\centering
	\begin{tabular}{|c|c|c|c|c|c|c|c|}
		\hline
		Problem size & (7,2) & (5,3) & (8,2) & (10,2)\\
		\hline
		Ogap   &  2.91\%& 10.59\% &  2.64\%& 2.09\%\\
		\hline
		Speed   & 3.39x & 6.74x  &  7.90x&   2.46x\\
		\hline
		\tabincell{c}{Optimal recognition\\ rate}    & 91.55\% & 91.55\% &  92.50\%& 95.35\% \\
		\hline
		Extra prune rate    & 24.09\% &  42.45\%& 31.58\% &  3.9\%\\
		\hline
	\end{tabular}
\end{table}

\subsection{Soft-Decision Algorithm}
We have thus far used SVM to train the classifier to learn the optimal prune policy. However, it cannot achieve dynamic control over optimality and computational complexity. Because the result of SVM classifier is either 0 or 1. Once the classifier is decided, the category of each node is determined and we cannot modify any more. If we want to get a less optimal solution with higher speed, we have to turn to a new policy, which is not convenient in practical systems.

To deal with this problem, we adopt the forward neural network (FNN) to replace SVM as the classifier inspired by \cite{nn,LORA}. For the binary classification problem, FNN can output the probability of each class rather than the definite result. It provides us with a soft-decision result and we can use it to dynamically control the optimality and computational complexity.

Specifically, we construct a $J$-layer FNN. We use Relu, i.e., Relu($x$) = max($0,x$), as the activation function for $J-1$ hidden layers and use softmax for the output layer to indicate the probability of each class. 	The label $\bm{y}$ of the optimal and non-optimal nodes are denoted as $(1,0)$ and $(0,1)$, respectively. And the loss function is the weighted cross-entropy,
as $$Loss = -\bm{\omega}[1]\bm{y}[1]\log(\bm{o}[1])-\bm{\omega}[2]\bm{y}[2]\log(\bm{o}[2]),$$
where $\bm{o}$ is the output vector of the FNN and $\bm{\omega}$ is the weight vector for each class. $\bm{\omega}$  is closely related to $\omega_1$ and $\omega_2$ mentioned in Section IV-C. Specifically, $\bm{\omega}[1] = \omega_1 \times \omega_2$ and $\bm{\omega}[2] = \omega_1$. This loss function is different from the one in \cite{LORA} that does not take the depth-dependent weights, $\omega_1$,  into consideration.

With the help of DAgger, FNN can be trained in a similar way to SVM classifier suggested in Tables II and III. After FNN-based policy is learned, we can get the output probability vector, $\bm{o}_n$, for node $n$, during the B\&B searching process. Then we compare the optimal probability $\bm{o}_n[1]$ with a given threshold, $\tau$. If $\bm{o}_n[1]>\tau$,  node $n$ is classified as the optimal node and is to be branched on; otherwise, node $n$ is classified as the non-optimal node and is to be pruned. The threshold, $\tau$, can be used to control the optimality and computational complexity trade-off. For example, if we want to get a less optimal solution with higher speed, we do not need to train a new policy. We only need to reduce $\tau$ iteratively to satisfy our goal. This dynamic control process is summarized in Table X.

\begin{table}[h]
	\setlength{\abovecaptionskip}{-2pt}
	\setlength{\belowcaptionskip}{-6pt}
	\caption{Dynamic Soft-Decision Algorithm}
	\label{A4}
	\begin{algorithm}[H]
		\caption{Dynamic Soft-Decision Algorithm}
		{\normalsize
	\begin{algorithmic}[1]
		\State \textbf{initialization}
		\begin{itemize}
			\item Set node list: $\mathscr{N}_Q\leftarrow\{n([ ])\}$.
			\item Set threshold: $\tau = 0.5$.
		\end{itemize}
		\While{\emph{Speed} \bm{or} \emph{Ogap} is not achieved}
		\While{$\mathscr{N}_Q \neq \varnothing$}
		\State \textbf{Node Selection:} pop the first node $n_Q(\bm{\rho}^d)$ in $\mathscr{N}_Q$.
		\State \textbf{Evaluation:} use interior-point method solve Problem (\ref{Mb}) corresponding to $n_Q(\bm{\rho}^d)$.
		\State \textbf{Fathom Decision:}
		\If {Node $n_Q(\bm{\rho}^d)$ is fathomed by original B\&B}
		\State Go to step 2.
		\Else
		\If {$\pi_p(n_Q(\bm{\rho}^d))[1] >= \tau$}
		\State Branch on node $n_Q(\bm{\rho}^d)$.
		\State \textbf{Variable Selection:} branch on the first undetermined variable in $\bm{\hat{\rho}}$.
		\State Add the new nodes into $\mathscr{N}_Q$.
		\EndIf
		\State Go to step 2.
		\EndIf
		\EndWhile
		\State Increase or Decrease $\tau$.
		\EndWhile
    \end{algorithmic}}
\end{algorithm}
\end{table}

\begin{table}
	\small
	\caption{Performance of SVM-based and FNN-based Imitation Learning \protect\\ for the Scenario with $K=5$, $L=2$}
	\label{table7}
	\centering
	\begin{tabular}{|c|c|}
		\hline
		Method & Speed\\
		\hline
		SVM-based imitation learning  & 2.06x\\
		\hline
		FNN-based imitation learning with new loss function&  1.74x\\
		\hline
		FNN-based imitation learning with loss function in \cite{LORA}&1.69x\\
		\hline
	\end{tabular}
	\end{table}
We use the scenario with $K=5$, $L=2$ as an example to compare the performance of the SVM-based and FNN-based imitation learning methods. We use 200 training samples and 20 testing samples for each method. In order to make a fair comparison with the loss function used in \cite{LORA}, we adopt the FNN structure in \cite{LORA} and construct an FNN with three hidden layers with 16, 32, and 16 nodes, respectively. We set the number of training epochs to be 30 and the batch size to be 128. We set the \emph{ogap} limit of the FNN-based method as the \emph{ogap} that can be attained by the SVM-based method, i.e., 2.01\%. We change $\tau$ by 0.01 for each iteration. The comparison results are summarized in Table \ref{table7}. Because the \emph{optimal recognition rate} and \emph{extra prune rate} change iteratively with $\tau$ in the FNN-based method, we only focus on \emph{speed} here to compare the SVM-based and FNN-based methods in a fair way. From Table \ref{table7}, the computational complexity of our proposed method is slightly lower than that in \cite{LORA}. It also suggests that the speed achieved by the FNN-based method is lower than that of the SVM-based method while achieving the same optimality. Because the FNN-based method needs iteration process to tune appropriate threshold $\tau$ for each problem, it will consume more time than the SVM-based method to achieve dynamic control.

%%% ------------------------- \input{sub/conclusion.tex} --------------
\section{Conclusions and Future Research}
This paper introduces a hybrid ML and mathematical method to solve resource allocation problems in D2D communication networks, which are usually formulated as MINLP problems.  The main goal of leveraging the ML technique is to overcome the disadvantages of traditional mathematical techniques and get an efficient algorithm for resource allocation in D2D communications. The key idea is to accelerate the B\&B algorithm, the widely-used globally optimal algorithm for MINLP problems, by learning a good auxiliary prune policy. This is achieved by the imitation learning method. Extensive experiment results have demonstrated that the proposed method can achieve good optimality and reduce computational complexity at the same time with only hundreds of training samples. It also has the generalization ability to large-scale problems. These two aforementioned properties are preferred in wireless networks. To further improve the generalization ability of the proposed method, we have also developed the mixed training strategy, which can balance the learned policy's performance and the cost of getting training samples from large-scale problems. Moreover, we have utilized a DNN as the binary classifier and proposed a novel loss function to dynamically control the optimality and computational complexity. Our studies in this work can be also applied to the MINLP problems in other wireless communication systems, such as communication mode selection and multi-cell resource allocation problems in D2D networks \cite{number2,liu1,liu2}.

There still exist some  problems about accelerating the B\&B algorithm for resource allocation problems by imitation learning method. First, the optimality of the proposed method is guaranteed by the optimality of the B\&B algorithm. However, optimal solutions can even not be achieved by the B\&B algorithm for more difficult problems whose corresponding nonlinear problems are non-convex. Our proposed method can still be used for the non-convex problems to achieve a faster B\&B algorithm, but the optimality cannot be guaranteed since the B\&B algorithm is not optimal in that case. Second, our proposed method can speed up the B\&B algorithm, but sometimes the speed up rates are not high enough as the test results suggest. Attempting to further speed up the B\&B algorithm while remaining the optimality at the same time is an important future direction. Also, finding other effective method to avoid dataset imbalance problem, further reducing the number of training samples, and even adopting new ML techniques to accelerate resource allocation are very interesting issues for fine-tuning our proposal.

%%% ---------------- \input{sub/appendix.tex} --------------------------

\begin{appendices}
\vspace{-0.3cm}
\section{Proof of Proposition 1}
From \cite{convex}, if D2D pair $l$ is allowed to reuse the channel of CU $k$, the following equation should be satisfied to maximize the minimum data rate of all D2D pairs and guarantee the data rate of CU $k$
$$\log(1+\frac{p_k^\text{C}h_{k}^\text{CB}}{\sigma_N^2+p_{kl}^\text{D}h_{l}^\text{DB}})=R_{\min}^\text{C}.$$
Then, we have
$$p_k^\text{C}=(2^{R_{\min}^\text{C}}-1)\frac{\sigma_N^2+p_{kl}^\text{D}h_l^{\text{DB}}}{h_k^\text{CB}}.$$
From (\ref{resource_sub3}) and (\ref{resource_sub5}), we can get the new bound of $p_{kl}^\text{D}$ as $p_{kl}^\text{D}\leq p_{kl}^{\max},$ where $$p_{kl}^{\max}=\min\{P_{\max}^\text{D},(1/h_l^\text{DB})(p_{\max}^\text{C}h_k^\text{CB}/(2^{R_{\min}^\text{C}}-1)-\sigma_N^2)\}.$$

\vspace{-0.2em}
\section{Proof of Proposition 2}
From Proposition 1, the data rate of D2D pair $l$ on the channel of CU $k$ can be rewritten as
\begin{small}
	\begin{eqnarray*}
		R_{kl}^\text{D}(\bm{p^\text{C}},\bm{p^\text{D}},\bm{\rho})&=&\log(1+\frac{\rho_{kl}p_{kl}^\text{D}h_l^\text{D}}{\sigma_N^2+p_k^\text{C}h_{kl}^\text{CD}})\\ &=&\log(1+\frac{\rho_{kl}p_{kl}^\text{D}h_l^\text{D}}{\sigma_N^2+\frac{(2^{R_{\min}^\text{C}}-1)h_{kl}^\text{CD}}{h_k^\text{CB}}(\sigma_N^2+p_{kl}^\text{D}h_l^\text{DB})})\\
		&\triangleq& \hat{R}_{kl}^\text{D}(\bm{p^\text{D}},\bm{\rho}).
	\end{eqnarray*}
\end{small}For convenience, we introduce
$$a_{kl}\triangleq \frac{\sigma_N^2}{h_l^\text{D}}+\frac{(2^{R_{\min}^\text{C}}-1)h_{kl}^\text{CD}\sigma_N^2}{h_l^\text{D}h_k^\text{CB}},$$
\vspace{-1em}
$$b_{kl}\triangleq \frac{(2^{R_{\min}^\text{C}}-1)h_{kl}^\text{CD}h_l^\text{DB}}{h_l^\text{D}h_k^\text{CB}}.$$
Then, $\hat{R}_{kl}^\text{D}(\bm{p^\text{D}},\bm{\rho})$ can be rewritten as
\begin{align}\label{abr}
\hat{R}_{kl}^\text{D}(\bm{p^\text{D}},\bm{\rho})=\log(1+\frac{\rho_{kl}p_{kl}^\text{D}}{a_{kl}+b_{kl}p_{kl}^\text{D}}).
\end{align}

Furthermore, we can rewrite (\ref{drate}) as
\begin{eqnarray*}
	R_{l}^\text{D}(\bm{p^\text{C}},\bm{p^\text{D}},\bm{\rho}) &=& \sum_{k\in \mathscr{K}}\rho_{kl} R_{kl}^\text{D}(\bm{p^\text{C}},\bm{p^\text{D}},\bm{\rho}) \\
	&=& \sum_{k\in \mathscr{K}}\rho_{kl} \log(1+\frac{\rho_{kl}p_{kl}^\text{D}h_l^\text{D}}{\sigma_N^2+p_k^\text{C}h_{kl}^\text{CD}})\\
	&=&\sum_{k\in\mathscr{K}}\log(1+\frac{\rho_{kl}p_{kl}^\text{D}h_l^\text{D}}{\sigma_N^2+p_k^\text{C}h_{kl}^\text{CD}})\\
	&=&\sum_{k\in \mathscr{K}}\log(1+\frac{\rho_{kl}p_{kl}^\text{D}}{a_{kl}+b_{kl}p_{kl}^\text{D}})\\
	&\triangleq& \hat{R}_{l}^\text{D}(\bm{p^\text{D}},\bm{\rho}).
\end{eqnarray*}
This equation always holds because  $\rho_{kl}$ is a binary variable.
\end{appendices}

\end{document}